\documentclass[preprint,12pt,authoryear,review]{elsarticle}

\usepackage{amsmath,amssymb,lineno}
\usepackage{amssymb}
\usepackage{soul}

\journal{Icarus}

\begin{document}

\begin{frontmatter}

\title{The stability of tightly-packed, evenly-spaced systems of Earth-mass planets orbiting a Sun-like star}

\author[DAA]{Alysa Obertas\corref{mycorrespondingauthor}}
\cortext[mycorrespondingauthor]{Corresponding author. Tel.: +1 416-978-7104}
\ead{obertas@astro.utoronto.ca}
\author[CITA]{Christa Van Laerhoven}
\author[CITA,CPS]{Daniel Tamayo}

\address[DAA]{Department of Astronomy \& Astrophysics, University of Toronto, 50 St George St., Toronto, ON, M5S 3H4}
\address[CITA]{Canadian Institute for Theoretical Astrophysics, 60 St George St., Toronto, ON, M5S 3H8}
\address[CPS]{Centre for Planetary Sciences, University of Toronto - Scarborough, 1265 Military Trail Office SW 504E, Toronto, ON, M1C 1A4}

\begin{abstract}
Many of the multi-planet systems discovered to date have been notable for their compactness, with neighbouring planets closer together than any in the Solar System. Interestingly, planet-hosting stars have a wide range of ages, suggesting that such compact systems can survive for extended periods of time. We have used numerical simulations to investigate how quickly systems go unstable in relation to the spacing between planets, focusing on hypothetical systems of Earth-mass planets on evenly-spaced orbits (in mutual Hill radii). In general, the further apart the planets are initially, the longer it takes for a pair of planets to undergo a close encounter. We recover the results of previous studies, showing a linear trend in the initial planet spacing between 3 and 8 mutual Hill radii and the logarithm of the stability time. Investigating thousands of simulations with spacings up to 13 mutual Hill radii reveals distinct modulations superimposed on this relationship in the vicinity of first and second-order mean motion resonances of adjacent and next-adjacent planets. We discuss the impact of this structure and the implications on the stability of compact multi-planet systems. Applying the outcomes of our simulations, we show that isolated systems of up to five Earth-mass planets can fit in the habitable zone of a Sun-like star without close encounters for at least $10^9$ orbits. 
\end{abstract}

\begin{keyword}

Planetary dynamics \sep Celestial mechanics \sep Extra-solar planets

\end{keyword}

\end{frontmatter}

% \linenumbers

%%%%%%%%%%%%%%%%%%%%%%%%%%%%%%%%%%%%%%%%%%%%%%%%%%%%%%%%%%%%%%%%%%%%%%%%%%%%%%%%%%%%%%%%%%%%%%%%%%%
%%%%%%%%%%%%%%%%%%%%%%%%%%%%%%%%%%%%%%%%%%%%%%%%%%%%%%%%%%%%%%%%%%%%%%%%%%%%%%%%%%%%%%%%%%%%%%%%%%%

\section{Introduction}
\label{Introduction}

Since the first detection of exoplanets over 20 years ago, more than 3500 exoplanets in over 2600\footnote{exoplanets.eu, retrieved November 28, 2016} systems have been discovered. Nearly 600 of these are multi-planet systems and are very different from the Solar System. These systems provide a testbed to investigate the effects of dynamics on planetary systems, from their formation to their long-term evolution. 

The Kepler mission has detected over 400 multi-planet systems and some of the most remarkable of these are compact systems with planets in close orbits. Kepler-11 is the archetype of these types of systems, having 5 planets within Mercury's orbit and a sixth just beyond \citep{Lis11}. Similar systems have been discovered since, including Kepler-33 \citep{Lis12}, Kepler-32 \citep{Swi13}, Kepler-80 \citep{Mac16}, and Kepler-444 \citep{Cam15}. As Kepler systems, they are generally not young (typical ages of Kepler stars are $>1~\mathrm{Gyr}$, \citet{Wal13,Mar14}) and are even as old as $11~\mathrm{Gyr}$ in the case of Kepler-444 \citep{Cam15}. In addition to Kepler systems, other missions have discovered compact multi-planet systems. An interesting example is the TRAPPIST-1 system which contains seven Earth-size planets, several of which orbit within the habitable zone \citep{Gil17}.

There have been numerous investigations of the stability of these specific systems (e.g. Kepler-11: \citet{Lis13}, \citet{Mah14}, and several others), which examine the outcomes of long-term integrations, but these systems also inspire continued theoretical study of the dynamical stability of tightly packed systems in general. 

For a two-planet system, there is an analytic criterion for the critical separation beyond which the system is Hill stable and the planets will never undergo a close encounter \citep{Mar82,Gla93}. Overlap of first-order mean motion resonances in two-planet systems yields another analytic criterion for a critical separation \citep{Wis80,Dec13}. No such analytic criterion has been found if a third planet is added to the system. Most results on stability timescales for arbitrary systems with three or more planets therefore come from  N-body integrations. Adding another planet to a system breaks the two-planet stability criterion. In general, such studies find that the time it takes for a close encounter to occur in such a system increases roughly exponentially with the initial spacing of the planets, albeit with large scatter in stability times about fitted trends. (\citet{Cha96,Mar02,Mor16}, among others). 

Curiously, based on a limited set of integrations, \citet{Smi09} observed a dramatic increase in the stability time in equally spaced five-planet systems with initial separations greater than $\sim 8$ mutual Hill radii, reaching $\sim 10$ Gyr at approximately 9 mutual Hill radii. \citet{Qui11} suggested that this transition marks the end of the region in which three-body resonances overlap, leading to increased stability at large separations.

Is the scatter in stability times intrinsic and just the outcome of chaotic interactions, or is there underlying structure? What changes as planets are spaced further apart, allowing for survival over billions of orbits? Motivated by these studies, we have simulated thousands of equal-mass and tightly packed systems of five planets spaced evenly in mutual Hill radii. We examine the relationship between the time until a close encounter between a pair of planets and the interplanetary spacing. 

We present an overview of previous studies in Section~\ref{s:back}, leading into a description of our methodology in Section~\ref{s:method}. We describe our results in Section~\ref{s:res} with discussion of the implications in Section~\ref{s:dis} and comparison of our results to previous studies. Our conclusions and recommendations for future studies are presented in Section~\ref{s:con}.

%%%%%%%%%%%%%%%%%%%%%%%%%%%%%%%%%%%%%%%%%%%%%%%%%%%%%%%%%%%%%%%%%%%%%%%%%%%%%%%%%%%%%%%%%%%%%%%%%%%
%%%%%%%%%%%%%%%%%%%%%%%%%%%%%%%%%%%%%%%%%%%%%%%%%%%%%%%%%%%%%%%%%%%%%%%%%%%%%%%%%%%%%%%%%%%%%%%%%%%

\section{Multi-planet Systems}
\label{s:back}

\subsection{Two-planet Systems}
\label{ss:two}

The Hill stability of the general three-body problem was investigated by \citet{Mar82}. \citet{Gla93} applied their results to the specific case of two planets orbiting a much more massive star. By expanding in powers of the planet-star mass ratio in the circular and co-planar limit, \citet{Gla93} showed that close encounters are forbidden if the outer planet has an initial semi-major axis ($a$),

\begin{equation}
    a_2 > a_1 (1 + \widetilde{\Delta}_C),
\end{equation}
where the subscripts 1 and 2 denote the inner and outer planet, respectively, and $\widetilde{\Delta}_C = 2 \cdot 3^{1/6} (\mu_1 + \mu_2)^{1/3}$ is the critical separation, to lowest order in the planet-to-star mass ratio $\mu_i = m_i/M$.

In terms of the mutual Hill radius $R_{H_{1,2}}$, this criterion is expressed as

\begin{equation}
    a_2 > a_1 + \Delta_C R_{H_{1,2}},
\end{equation}
where the mutual Hill radius between adjacent bodies is
\begin{equation}
\label{eq:hill}
    R_{H_{i,i+1}} =  \left( \frac{\mu_i + \mu_{i+1}}{3} \right)^{1/3} \left(\frac{a_i + a_{i+1}}{2}\right).
\end{equation}

Defining
\begin{equation}
\label{eq:X}
    X = \frac{1}{2} \left( \frac{\mu_i + \mu_{i+1}}{3}, \right)^{1/3}
\end{equation}
the critical two-planet separation in units of the mutual Hill radius is then
\begin{equation}
\label{eq:crit-sep}
    \Delta_C = \frac{2\sqrt{3}}{1+2\sqrt{3}X}
\end{equation}

In simulations of two-planet systems, \citet{Gla93} found that those with an initial separation less than this critical value had close encounters between the two planets in relatively few conjunctions (up to $\sim 10^2$), even for separations up to $\sim 1$\% below the critical value.

An alternative approach is to derive the critical separation at which adjacent first-order mean motion resonances (MMRs) overlap, driving widespread chaos \citep{Wis80}. To first order in the planets' masses, eccentricities, and inclinations, \citet{Dec13} show that first-order MMRs overlap if the planets satisfy

\begin{equation}
\label{eq:mmr-overlap}
    \frac{a_2-a_1}{a_1} \lesssim 1.46(\mu_1+\mu_2)^{2/7}
\end{equation}

These two criteria therefore scale very similarly with planetary masses, and cross at $(\mu_1+\mu_2)\simeq 3\times 10^{-5}$ (i.e. combined planet mass of $\sim 10 M_{\oplus}$ around a solar-mass star). For larger masses, the critical resonance overlap separation becomes smaller than the critical close encounter separation.

%%%%%%%%%%%%%%%%%%%%%%%%%%%%%%%%%%%%%%%%%%%%%%%%%%%%%%%%%%%%%%%%%%%%%%%%%%%%%%%%%%%%%%%%%%%%%%%%%%%

\subsection{Systems Of More Than Two Planets}
\label{ss:three}

Numerical investigations of the stability of multi-planet ($N\geq 3$) systems of planets in close orbits have been done \citep[e.g.,][]{Cha96,Mar02,Smi09,Pu15,Tam15,Mor16}, motivated by the consequences for planetesimal disks, closely-spaced multi-planet systems, and planet-induced gaps in debris disks. The same dynamics are relevant for Uranus's inner satellites, with stability timescales orders of magnitude shorter than the age of the Solar System \citep{Dun97,Fre12}. Destined for close encounters, they have perhaps collided and re-formed throughout the history of the Solar System \citep{Fre12}.

\citet{Pu15} propose that systems generally form with many planets on closely-spaced orbits, but those within stability limits have already gone unstable, only leaving the stable multi-planet systems in existence. Indeed, the Solar System may have had its own system of tightly packed planets interior to Venus's orbit that went unstable and led to collisions, leaving Mercury behind from the debris \citep{Vol15}.

\citet{Cha96} found that the logarithm of the time it takes for systems of equally spaced planets (in mutual Hill radii) to go unstable increases linearly with the initial interplanetary spacing. This finding has been confirmed by several follow-up studies \citep[e.g.,][]{Fab07,Smi09,Mor16}. There are slight variations in how different studies define the initial separation between planets (e.g. by using a different definition of the mutual Hill radius) and the stability criterion (either a close encounter or when orbits cross), but they show the same relationship in stability time $t_c$ relative to the initial period $t_0$ of the inner planet for initial separations (in units of mutual Hill radii) in the range $\Delta_C \lesssim \Delta \lesssim 8$,

\begin{equation}
\label{eq:logt}
    \log{(t_c/t_0)} = b \Delta + c
\end{equation}

Comparing systems of three and five planets, for the same initial separation it takes less time in a five planet system for a close encounter to occur between a pair of planets \citep{Cha96}. They also showed that increasing the number of planets further does not show significant change in the close encounter time, which suggest that only the closest neighbouring planets perturb each other.

Additional works have studied the effects of additional giant planets beyond the tightly spaced group. The difference such bodies make varies widely with their level of dynamical excitation, and with their separation from the compact system. In N-body simulations of low-mass protoplanets, \citet{Ito99} found that Jupiter-mass planets can substantially decrease the stability time, and \cite{Huang16} found similar trends with outer giant planets undergoing dynamical instabilities. By contrast, \citet{Smi09} found that a more widely spaced Jupiter-mass planet had smaller effects on systems of five Earth-mass planets, and \citet{Dun97} saw no change in orbit crossing times between sets of integrations with and without the large regular moons in the Uranian system.

%%%%%%%%%%%%%%%%%%%%%%%%%%%%%%%%%%%%%%%%%%%%%%%%%%%%%%%%%%%%%%%%%%%%%%%%%%%%%%%%%%%%%%%%%%%%%%%%%%%
%%%%%%%%%%%%%%%%%%%%%%%%%%%%%%%%%%%%%%%%%%%%%%%%%%%%%%%%%%%%%%%%%%%%%%%%%%%%%%%%%%%%%%%%%%%%%%%%%%%

\section{Methods}
\label{s:method}

In this investigation, we consider systems of five Earth-mass planets evenly spaced in mutual Hill radii with co-planar and initially circular orbits around a solar-mass star. 

The initial semi-major axis of the inner planet is fixed at $a_1=0.99AU$ for all systems. The logic for this choice is discussed in Section \ref{ss:hab}. The time can be non-dimensionalised by the innermost planet's orbital period, but choice of a particular $a_1$ allows number of orbits to be expressed as a length of time. The semi-major axes of the remaining planets are at fixed intervals in units of the mutual Hill radius (eq.~\eqref{eq:hill}),

\begin{equation}
\label{eq:anext}
    a_{i+1} = a_i + \Delta R_{H_{i,i+1}}
\end{equation}

The semi-major axis of the $n$-th planet relative to the $i$-th planet can be expressed in terms of the planet separation $\Delta$ and the parameter $X$\footnote{The reader should be aware that different groups call variables by different symbols. We use $\Delta$ for the number of mutual Hill Radii by which planets are separated, as does \citet{Cha96}. \citet{Smi09} instead use $\beta$. Our $X$ differs from $K$ in \citet{Cha96} by a factor of 2.} (defined in eq.~\eqref{eq:X}), which has a constant value for systems of equal-mass planets: 

\begin{equation}
\label{eq:sep}
    a_n = a_i \left( \frac{1+\Delta X}{1-\Delta X} \right) ^ {n-i}
\end{equation}

\noindent For Earth-mass planets, $X = 0.0063$ which results in $\Delta_C \simeq 3.4$ by eq.~\eqref{eq:crit-sep}.

A total of 17 500 five-planet systems with different initial conditions were generated. 16 000 values of $\Delta$ were drawn from a uniform distribution in [2.0, 10.0) and 1500 were drawn from a uniform distribution in [10.0, 13.0) (i.e. 25\% of the resolution for smaller $\Delta$). The initial phases of the planets were drawn uniformly on the circle, with angles in [0, $2\pi$). Comparing these systems to a set with a minimum angular separation imposed (to avoid initially placing planets near conjunction), the ensemble results did not differ for systems with $\Delta > \Delta_C$. 

The systems were integrated with the REBOUND N-body code \citep{Rei12} using the symplectic Wisdom-Holman integrator WHFast \citep{Rei15}. We adopted a timestep of $dt = 0.05$yr (i.e. the inner-most planet completes an orbit in approximately 20 timesteps) and integrations continued until a pair of planets had a close encounter or until 10Gyr elapsed. The time that the simulation terminated is defined as the close encounter time $t_c$ and a close encounter was defined as when the distance between any pair of planets became less than the Hill radius of the inner planet, $R_H = a_1(\mu_1/3)^{1/3}\sim 0.01AU$. Previous studies (e.g., \citet{Gla93}) have found that the exact stopping condition adopted is not important and we refer to $t_c$ as the stability time of a system. At circular velocities, the relative distance travelled between neighbouring planets in one timestep varies between $\sim 0.4 - 1.9R_H$. The relative distance increases with $\Delta$ and is $R_H$ at $\Delta=5.1$ and is $2R_H$ at $\Delta=10.3$. To check whether missing close encounters could significantly bias our results, we simulated systems with $4 \leq \Delta < 8$ using a wider close encounter distance of $2R_H$, and obtained indistinguishable results. Additionally, comparisons of ensemble close encounter times between different sets of simulations with $dt=0.05$yr and smaller time steps down to $dt=0.0005$ for spacings up to $\Delta = 7$ showed no discernible differences between the sets. The use of $dt=0.05$yr allowed integrations up to 10 billion orbits within reasonable computing time.

%%%%%%%%%%%%%%%%%%%%%%%%%%%%%%%%%%%%%%%%%%%%%%%%%%%%%%%%%%%%%%%%%%%%%%%%%%%%%%%%%%%%%%%%%%%%%%%%%%%
%%%%%%%%%%%%%%%%%%%%%%%%%%%%%%%%%%%%%%%%%%%%%%%%%%%%%%%%%%%%%%%%%%%%%%%%%%%%%%%%%%%%%%%%%%%%%%%%%%%

\section{Results}
\label{s:res}

In general, our results recover the trends for $\Delta_C \lesssim \Delta \lesssim 8$ found in previous investigations of the stability of multi-planet systems in close orbits. However, our much larger number of randomly spaced simulations reveal surprisingly regular variations in the stability timescales.

\begin{figure}
\begin{center}
    \includegraphics[width=\columnwidth,angle=90]{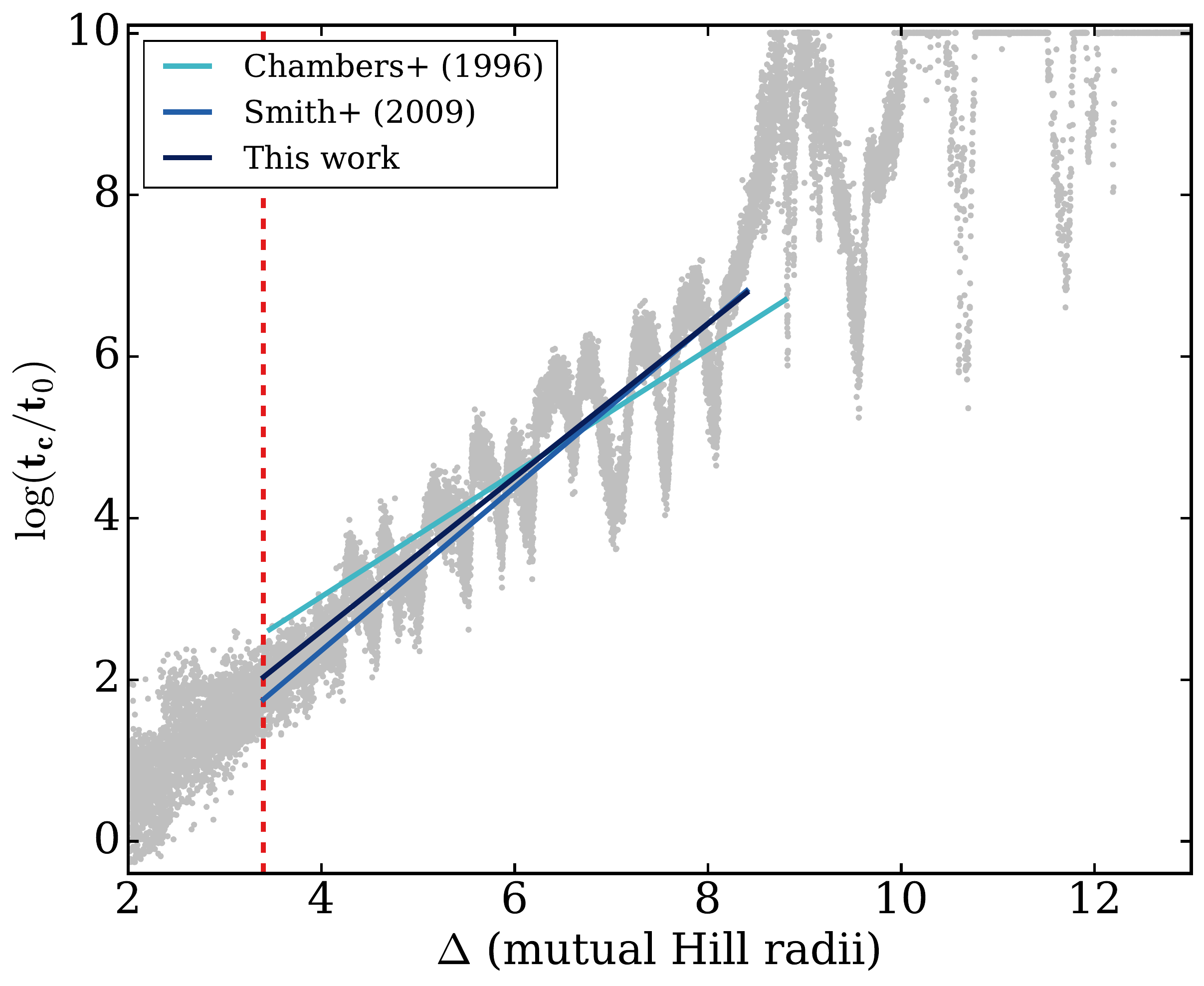}
    \caption[full data and fits]{Stability times $t_c$ of 17 500 simulations of five Earth-mass planet systems as a function of the initial separation between planets in mutual Hill radii (eq.~\eqref{eq:hill}). $t_0$ is the initial period of the inner planet. The maximum integration time was $\sim 10^{10}$ orbits. The vertical dashed line marks $\Delta = \Delta_C $ (eq.~\eqref{eq:crit-sep}, for $\mu_i= M_{\oplus}/M_{\odot}$). Fits to the stability times are shown for this work, \citet{Cha96}, and \citet{Smi09} (see Table~\ref{tab:lin-fit} for more details; note that the line for \citet{Cha96} is for a smaller $\mu_i$ than our work and \citet{Smi09}).}\label{fig:logt-delta}
  \end{center}
\end{figure}

The stability time (relative to the initial orbital period of the inner planet) as a function of the initial separation in Hill radii is shown in Figure~\ref{fig:logt-delta}. For $\Delta \lesssim \Delta_C$ (i.e. every pair of adjacent planets is initially within the critical two-planet separation for Hill stability) almost all systems have a close encounter within $\sim 100$ orbits. There is an overall trend of increasing stability time as the initial separation increases. There are distinct modulations superimposed on this trend, which start to appear for $\Delta \gtrsim \Delta_C$. The depths of these range from one to four orders of magnitude. For example, between $7.5 \lesssim \Delta \lesssim 7.75$, systems go from having close encounters within thousands of orbits to millions of orbits. Some features are sharp transitions in stability time (e.g. at $\Delta \simeq 5.5$) and some are more gradual dips (e.g.  at $\Delta \simeq 7$). 

Additionally, we confirm the claim of \citet{Smi09} that beyond $\Delta \gtrsim 8.4$, there is a sharp increase in the stability time, with many systems surviving for at least 10 billion orbits without close encounters. However, the limited number of wide-separation integrations by \citet{Smi09} did not capture either the substantial sharp drops in stability time visible in Fig.~\ref{fig:logt-delta} between $8 < \Delta < 9$. \citet{Smi09} only examined systems of five Earth-mass planets up to $\Delta = 8.8$. We include systems up to $\Delta < 13$, although most systems beyond $\Delta \sim 10$ have $\log(t_c/t_0) >= 9.95$. There is a broad, deep decrease near $\Delta \sim 9.5$ and two additional decreases near $\Delta \sim 10.6$ and $\Delta \sim 11.8$. The resolution of systems for $\Delta \geq 10$ is 25\% of the resolution for smaller values of $\Delta$, however.

A linear least-squares fit the logarithm of the stability times for initial separations $\Delta_C < \Delta < 8.4$ is also shown in Figure~\ref{fig:logt-delta}, in addition to the linear fits for five planet systems from \citet{Cha96} ($\mu_i=10^{-7}$) and \citet{Smi09} ($m_i = M_{\oplus}$, $M = 1 M_{\odot}$). The coefficients for these fits are shown in Table~\ref{tab:lin-fit} along with coefficients for initial separations $\Delta_C < \Delta < 10$ and $2.5 < \Delta < 8.4$.

The slope of the fit obtained by \citet{Cha96} is lower than the other values of $b$ shown in Table~\ref{tab:lin-fit}, although this is not surprising given the lower planet mass ratio. Comparing systems of three planets with mass ratios $\mu_i = 10^{-9}, 10^{-7},$ and $10^{-5}$, \citet{Cha96} obtained different slopes for each fit. After scaling the physical planet separation by $\mu_i^{1/4}$, linear fits between their three sets of integrations yielded similar slopes, which has been found by other studies (e.g. \citet{Fab07,Mor16}). This corresponds to scaling the planet spacing in Hill radii by $\mu_i^{1/12}$. Rescaling $b=0.765$ for $\mu_i=10^{-7}$ to $\mu_i=M_{\oplus}/M_{\odot}$ Earth-mass planets gives $b'=1.016$ which is consistent with that of \citet{Smi09} and this work.

Considering the last three fits shown in Table~\ref{tab:lin-fit}, the linear fits differ depending on the range of $\Delta$ used to calculate the coefficients $b$ and $c$. The ranges of $\Delta_C < \Delta < 10$ and $2.5 < \Delta < 8.4$ were used to compare the coefficients when extending the $\Delta$ range on either side. The lowest limit $\Delta > 2.5$ is used as systems with smaller spacings typically had a close encounter within the synodic period of the inner two planets.

The slopes vary between $b=0.951$ and $b=1.086$ and the intercepts range from $c=-1.202$ and $c=-1.881$ which results in different close encounter times using eq.\eqref{eq:logt}. For example, at $\Delta=8$, these coefficients give $\log(t_c/t_0)$ between 6.4 and 6.8. In contrast, the close encounter times shown in Figure~\ref{fig:logt-delta} near $\Delta=8$ range from $\log(t_c/t_0)=4.7$ to $7.2$.

We note that a major result of this work is that using a simple power law to describe the stability time does not capture the variation from that fit. For these simulations with Earth-mass planets, calculating stability times using eq.~\eqref{eq:logt} can give values that are up to two orders of magnitude higher or lower than the simulation values. 

\begin{table}
\begin{center}
\caption{Linear fits to the log of the simulation end time (due to a close encounter) versus initial planet separation in mutual Hill radii (eq.~\eqref{eq:hill}) for systems of five equal mass planets. The first three linear fits are displayed in Figure~\ref{fig:logt-delta}.}
    \begin{tabular}{ccccc}    
        \hline 
        Reference & b & c & range & $\mu_i$ \\ \hline \\
        \citet{Cha96} & 0.765 & -0.030 & $2\sqrt{3} < \Delta \leq 8.8$ & $10^{-7}$ \\
        \citet{Smi09} & 1.012 & -1.686 & $3.4 \leq \Delta \leq 8.4 $ & $3.0035\times 10^{-6}$ \\
        This work & 1.086 & -1.881 & $ \Delta_C < \Delta < 10 $ & $3.0035 \times 10^{-6}$ \\
        This work & 0.951 & -1.202 & $ \Delta_C < \Delta < 8.4 $ & $3.0035 \times 10^{-6}$ \\
        This work & 0.964 & -1.289 & $ 2.5 < \Delta < 8.4 $ & $3.0035 \times 10^{-6}$ \\ \\
        \end{tabular}
    \label{tab:lin-fit}
\end{center}
\end{table}

%%%%%%%%%%%%%%%%%%%%%%%%%%%%%%%%%%%%%%%%%%%%%%%%%%%%%%%%%%%%%%%%%%%%%%%%%%%%%%%%%%%%%%%%%%%%%%%%%%%

\subsection{Period Ratios}

A consequence of the even initial spacing of planets (in mutual Hill radii) is that the initial period ratios of pairs of planets are the same. Using eq.~\eqref{eq:sep} and Kepler's third law, the period ratio of the $n$-th planet and the $i$-th planet is,

\begin{equation}
    \frac{P_n}{P_i} = \left( \frac{1 + \Delta X}{1 - \Delta X} \right) ^{3(n-i)/2}. \label{eq:pratio}
\end{equation}

\noindent For adjacent pairs (i.e. $n=i+1$), the exponent is $3/2$. 

\begin{figure}
\begin{center}
   \includegraphics[width=\columnwidth,angle=90]{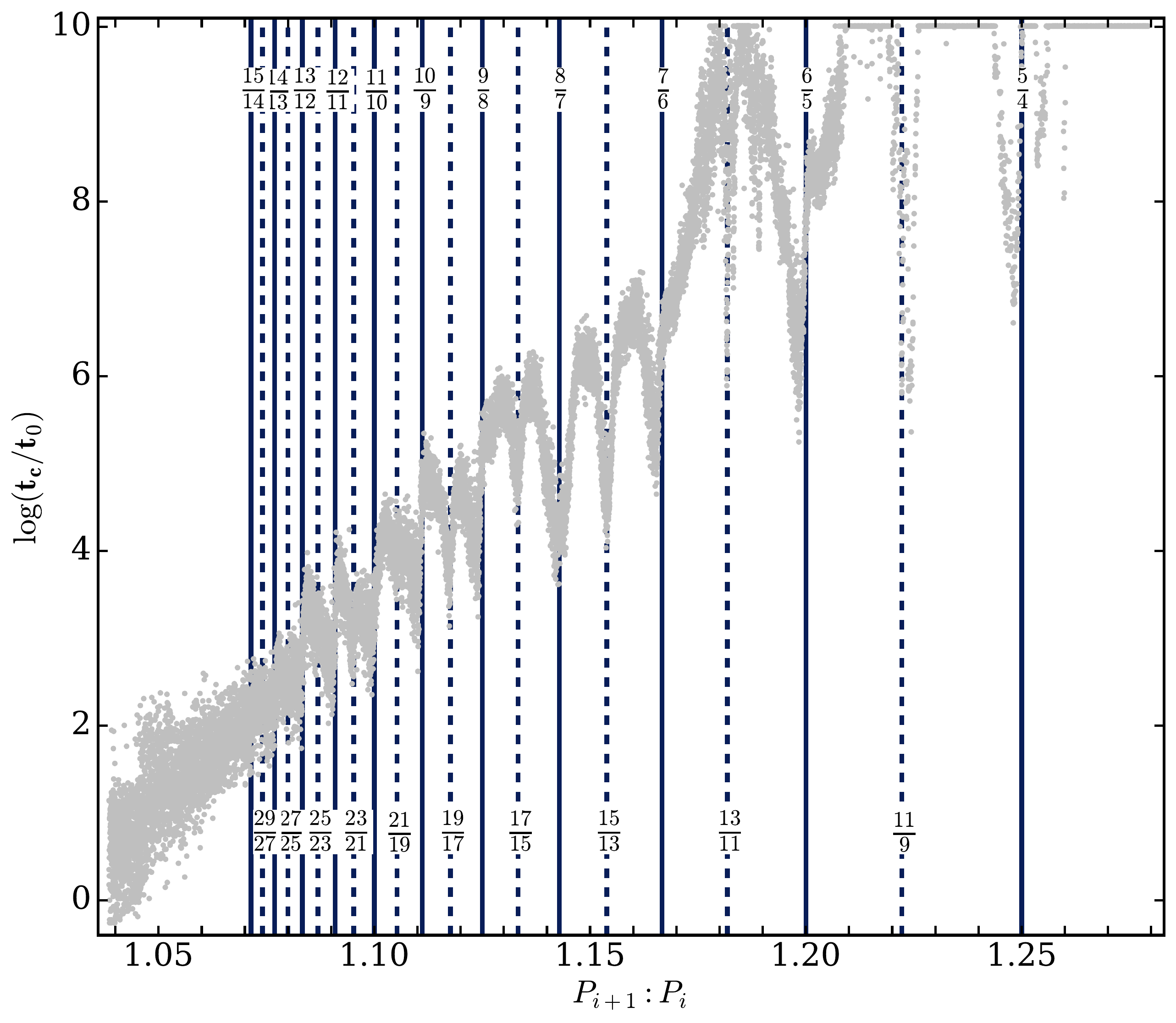}
   \caption[Period ratios adj]{Period ratios of all pairs of adjacent planets (eq.~\eqref{eq:pratio}). The vertical lines show the locations of different $(m+1):m$ (solid) and $(m+2):m$ (dashed) period ratios of adjacent planets, where pairs of planets are in the vicinity of first- and second-order MMRs.}\label{fig:pratio-adj}
 \end{center}
\end{figure}

\begin{figure}
\begin{center}
   \includegraphics[width=\columnwidth,angle=90]{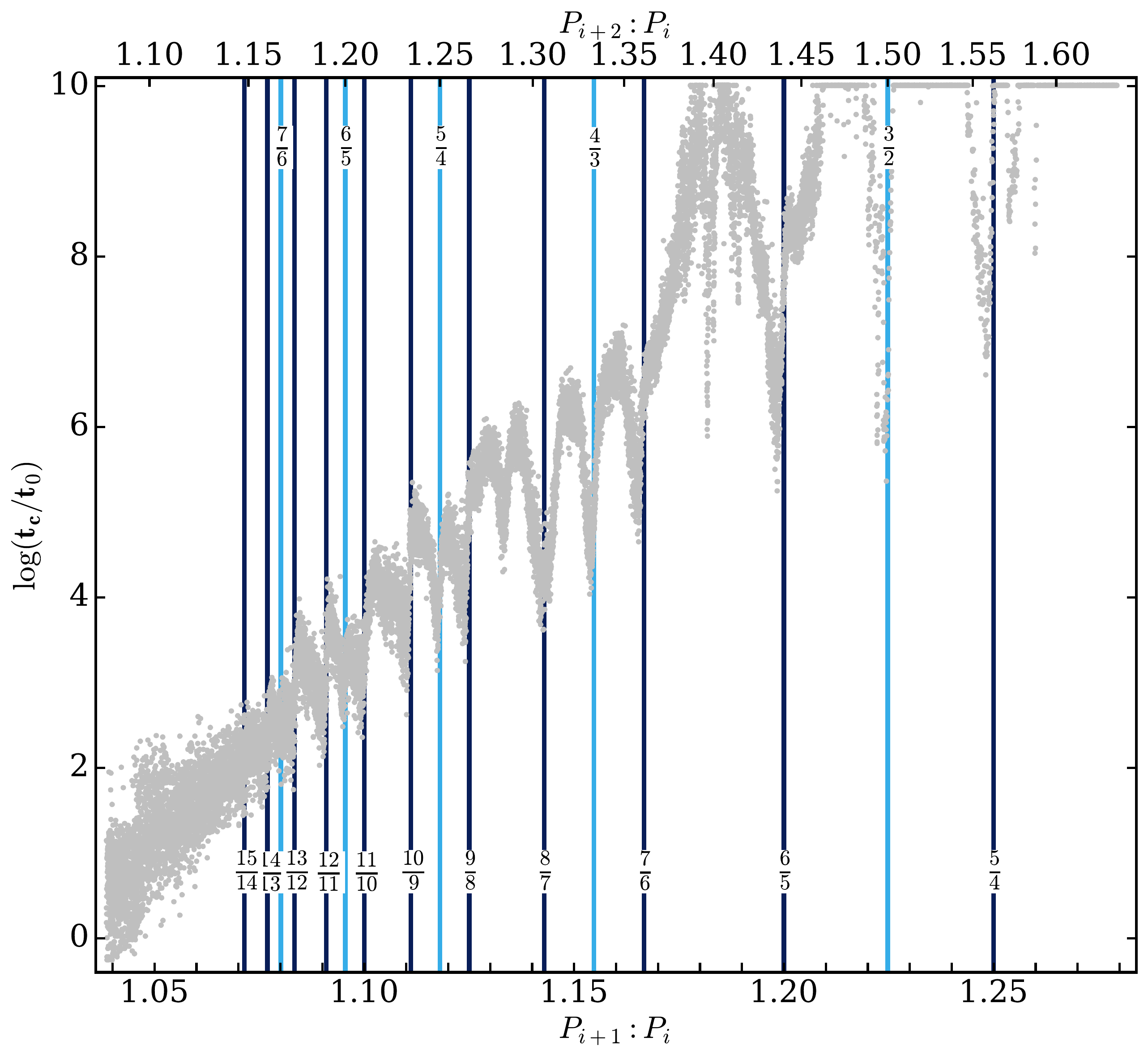}
   \caption[Period ratios next]{Period ratios of all pairs of adjacent planets (eq.~\eqref{eq:pratio}). The vertical lines show the locations of different $(m+1):m$ period ratios of adjacent planets (dark) and next-nearest (i.e. planets $i$ and $i+2$) planets (light), where pairs of planets are in the vicinity of first-order MMRs.}\label{fig:pratio-next}
 \end{center}
\end{figure}

The stability time as a function of the period ratio of adjacent pairs is shown in Figures~\ref{fig:pratio-adj} and \ref{fig:pratio-next}. Both include solid dark blue vertical lines corresponding to several $(m+1):m$ period ratios, which correspond to separations where pairs of adjacent planets are in the vicinity of first-order mean motion resonances (MMRs). Figure~\ref{fig:pratio-adj} also includes dashed dark blue vertical lines corresponding to $(m+2):m$ period ratios (i.e. in the vicinity of second-order MMRs) of pairs of adjacent planets. Figure~\ref{fig:pratio-next} also includes solid blue vertical lines corresponding to $(m+1):m$ period ratios of pairs of next-nearest planets (i.e. $n=i+2$ in eq.\eqref{eq:pratio}). As the planet spacing increases, the spacing between subsequent integer period ratios also increases. 

\begin{figure}
\begin{center}
   \includegraphics[width=\columnwidth]{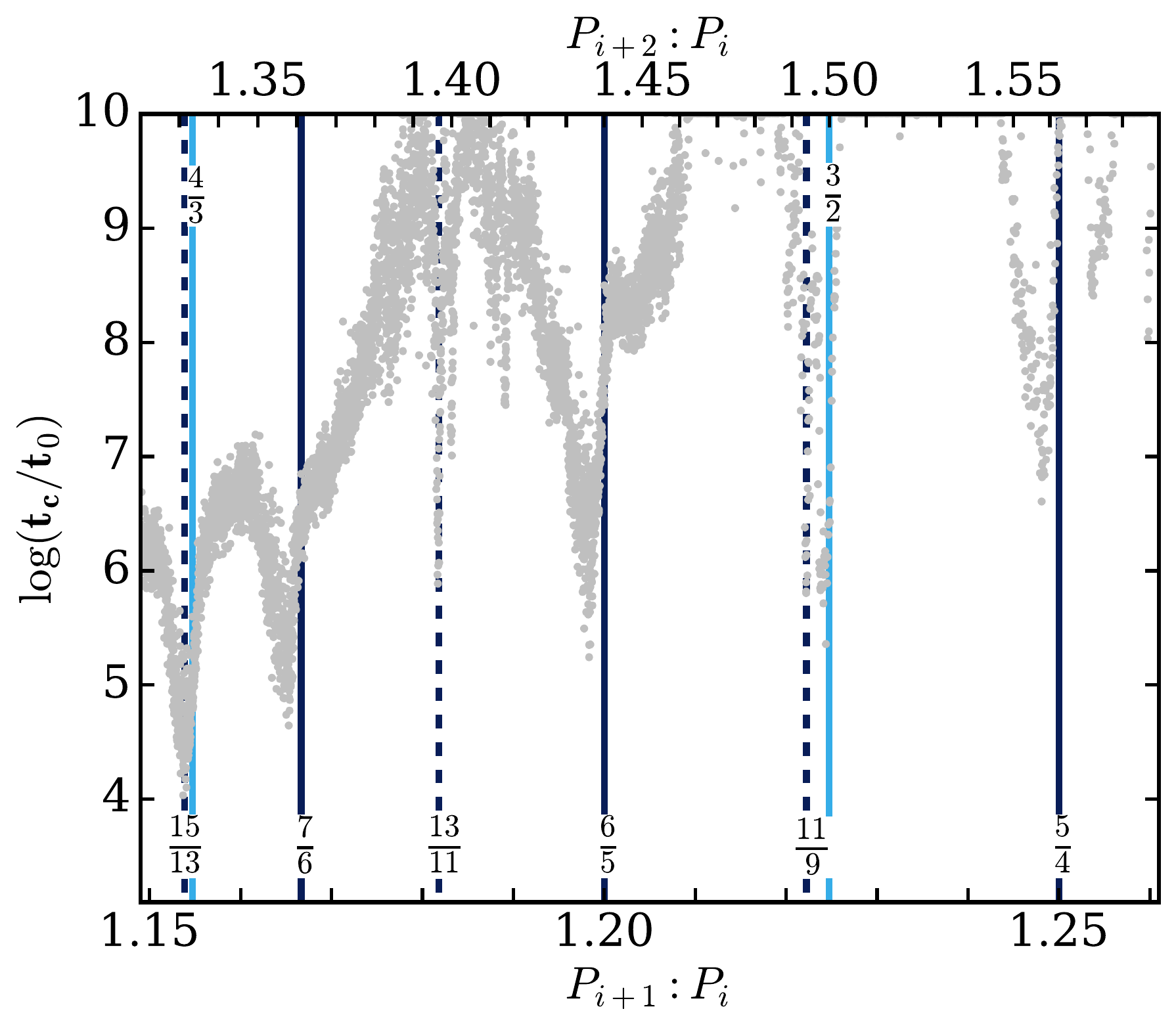}
   \caption[Period ratios]{Period ratios of all pairs of adjacent planets (eq.~\eqref{eq:pratio}) for $7.3 < \Delta < 12.4$.}\label{fig:zoom}
 \end{center}
\end{figure}

%%%%%%%%%%%%%%%%%%%%%%%%%%%%%%%%%%%%%%%%%%%%%%%%%%%%%%%%%%%%%%%%%%%%%%%%%%%%%%%%%%%%%%%%%%%%%%%%%%%
%%%%%%%%%%%%%%%%%%%%%%%%%%%%%%%%%%%%%%%%%%%%%%%%%%%%%%%%%%%%%%%%%%%%%%%%%%%%%%%%%%%%%%%%%%%%%%%%%%%

\section{Discussion}
\label{s:dis}

\subsection{Mean Motion Resonances}

The stability times show drastic changes in the vicinity of first and second-order mean motion resonances (MMRs) between adjacent and next-nearest planets. Local minima in $t_c/t_0$ are associated with the nominal period ratios of first and second-order MMRs, and we hypothesise that this is caused by overlap of strong MMRs driving chaotic diffusion \citep[e.g.,][]{Chirikov79, Murray99}.

Similar variations in the stability time were also seen in \citet{Cha96}, although with smaller amplitudes and larger widths. \citet{Cha96} suggest that these were due to MMRs because planet pairs share the same period ratios, although the relationship was not investigated further. In simulations with non-equal masses and both non-equal mass and non-even initial spacing, \citet{Cha96} see these features greatly diminished, but the stability timescales are approximately similar for the same spacing. \citet{Pu15} also saw strong declines in survivability at period ratios corresponding to MMRs for systems with varied mass and inter-planet spacing, suggesting that the variations seen in our systems may be present with similarly varied initial conditions.

As mentioned above, equal Hill-sphere separations between adjacent bodies imply equal period ratios between nearest planets. This means that when a body is near an MMR with an interior neighbour, it is by construction near the same MMR with its outer neighbour. This is thus a particularly active dynamical configuration and, indeed, unequal separations between planets yield different stability times \citep{Marzari14}.

Additionally, each MMR can force an equilibrium eccentricity \citep[e.g.][]{SSD}. Because we initialise systems with zero eccentricity, planet pairs will in many cases perform large-amplitude oscillations around these fixed points, and we suspect this will further promote instability. Finally, because our initial orbital phases are drawn from uniform distributions [$0$,$2\pi$), most planet pairs will not have resonant angles near equilibrium values, again leading to large-amplitude variations. One might expect that systems initialised with eccentricities and phases putting them near resonant island centers to be longer lived.

Other than the $8:7$ MMR between adjacent planets, the locations of first-order MMRs in figures~\ref{fig:pratio-adj} and \ref{fig:pratio-next} are located at slightly larger period ratios than the minimum stability times of the nearby features. In contrast, the locations of second order MMRs are located at the minima of the nearby features. We do not currently understand the reason for this asymmetry.

The equal period ratios between adjacent planets introduces a number of degeneracies. For example, $m+1:m$ first-order MMRs between next-adjacent pairs of planets always lie close to the $4m+3:4m+1$ second-order MMRs between adjacent planets (a consequence of eq.\eqref{eq:pratio}). This makes it difficult to say which one predominantly drives the dynamics. However, while all first-order resonances between next adjacent planets lie close to second-order resonances between adjacent planets, the inverse is not true. For example, the dip near the 17:15 MMR between pairs of adjacent planets does not match the location of a first-order resonance between next-nearest neighbouring planets.

Beyond the $7:6$ MMR we see the sharp increase in stability time, corresponding to $\Delta \gtrsim 8.4$. Figure~\ref{fig:zoom} highlights the stability times of systems at these period ratios.

%%%%%%%%%%%%%%%%%%%%%%%%%%%%%%%%%%%%%%%%%%%%%%%%%%%%%%%%%%%%%%%%%%%%%%%%%%%%%%%%%%%%%%%%%%%%%%%%%%%

\subsection{Long-term Survival of Systems}

The drastic increase in stability times for systems with initial spacings of $8.4 \lesssim \Delta \lesssim 8.8$ was also seen by \citet{Smi09}, although there were only 6 simulations in that regime. Our results from over 5000 systems with $\Delta > 8$ also show that systems with larger initial inter-planet spacing can survive for billions of orbits without close encounters. 

\begin{figure}
\begin{center}
   \includegraphics[width=\columnwidth]{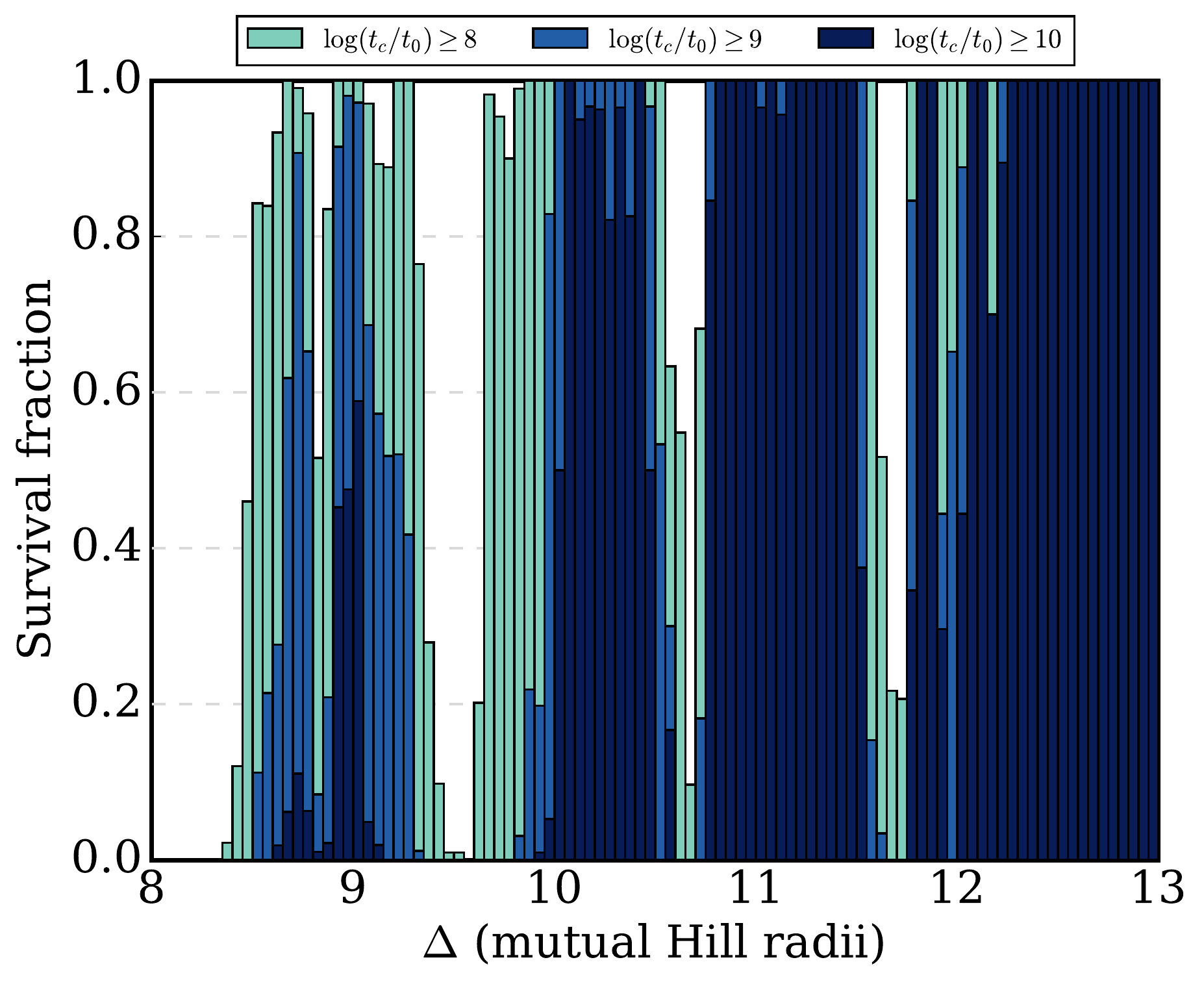}
   \caption[Survivability]{Fraction of systems in each bin with first close encounters after $10^8$, $10^9$, and $10^{10}$ orbits. Each bin has a width of 0.05. The mean total number of systems in each bin range is 100 for $8 \leq \Delta < 10$ and 25 for $10 \leq \Delta < 13$.}\label{fig:surv}
 \end{center}
\end{figure}

Figure~\ref{fig:surv} shows the fraction of systems within each histogram bin in $\Delta$ that have a first close encounter between a pair of planets within at least 100 million orbits (light blue), 1 billion orbits (blue), and 10 billion orbits (dark blue), which we refer to as the survival fraction for a given timescale. Each histogram bin has a width in $\Delta$ of 0.05. The mean total number of systems in each bin range is 100 for $8 \leq \Delta < 10$ and 25 for $10 \leq \Delta < 13$. Since the maximum integration time was 10Gyr ($\sim 10$ billion orbits), non-zero 10 billion orbit survival fractions represent systems that reached the maximum integration time without having any close encounters. The lower survival fractions around $\Delta \sim 8.8$, $\Delta \sim 9.6$, $\Delta \sim 10.6$, and $\Delta \sim 11.8$ are not surprising as these systems are in the vicinity of the $13:11$, $6:5$, $11:9$, and $5:4$ MMRs for adjacent planets.

A summary of the proportion of systems with stability times of at least $10^8$, $10^9$, and $10^{10}$ orbits for different ranges in $\Delta$ is shown in Table~\ref{tab:surv}. As evident in Figure~\ref{fig:surv}, most systems with spacings $8.4 \leq \Delta \leq 13$ survived for at least $10^8$ orbits without close encounters. Considering different ranges of $\Delta$, there are some differences in the proportions of systems that survive for at least $10^8$, $10^9$ and $10^{10}$, however.

\begin{table}
\begin{center}
\caption{Proportions of systems with stability times of at least $10^8$, $10^9$, and $10^{10}$ orbits in different ranges of $\Delta$.}
    \begin{tabular}{ccccc}    
        \\ \hline 
        Range & $N_{system}$ & $10^8$ & $10^9$ & $10^{10}$ \\ \hline \\
        $8.4 \leq \Delta \leq 13$ & 4690 & 79\% & 47\% & 28\% \\
        $8.4 \leq \Delta < 8.8$ & 814 & 76\% & 35\% & 3\% \\
        $8.8 \leq \Delta < 9.6$ & 1591 & 63\% & 37\% & 11\% \\
        $9.6 \leq \Delta < 10.6$ & 1099 & 90\% & 35\% & 20\% \\
        $10.6 \leq \Delta < 11.8$ & 628 & 83\% & 70\% & 64\% \\
        $11.8 \leq \Delta < 13$ & 558 & 100\% & 94\% & 89\% \\ \\
        \end{tabular}
    \label{tab:surv}
\end{center}
\end{table}

%%%%%%%%%%%%%%%%%%%%%%%%%%%%%%%%%%%%%%%%%%%%%%%%%%%%%%%%%%%%%%%%%%%%%%%%%%%%%%%%%%%%%%%%%%%%%%%%%%%

\subsection{Planets in the Habitable Zone}
\label{ss:hab}

For spacings up to $\Delta=10.7$, all five planets fit in the habitable zone as defined by \citet{Kop13}. Taking the moist greenhouse and maximum greenhouse limits calculated with 1D climate models, the inner edge of the habitable zone is at 0.99AU and the outer edge of the habitable zone is at 1.70AU. Our results imply that systems of five Earth-mass planets around a Solar-mass star with $8.6 < \Delta < 13$ could survive for at least 10 billion orbits. Placing the inner planet at $a_1 = 0.99AU$, eq.\eqref{eq:sep} gives $a_5 =1.53AU$ for $\Delta=8.6$. At this separation, a sixth planet would be just outside the outer edge initially with $a_6 > 1.70AU$. With $a_1=0.99AU$, such systems could have stability times exceeding the 10Gyr main-sequence lifetime of a Solar-mass star, although habitable zone boundaries would evolve over this time period.

A caveat to this conclusion is that the models used to calculate the boundaries of the habitable zone in \citet{Kop13} do not incorporate the physics of clouds and their effects on the climate. Warming and cooling due to clouds depend on properties such as their heights, particle size, and coverage, which means that 1D models that include clouds make simplifications that make it difficult to accurately calculate the habitable zone boundaries. Consequently, \citet{Kop13} suggest using 3D global circulation models. The use of these models is a developing area of research, but they will improve the calculation of exoplanet habitable zone boundaries.

%%%%%%%%%%%%%%%%%%%%%%%%%%%%%%%%%%%%%%%%%%%%%%%%%%%%%%%%%%%%%%%%%%%%%%%%%%%%%%%%%%%%%%%%%%%%%%%%%%%
%%%%%%%%%%%%%%%%%%%%%%%%%%%%%%%%%%%%%%%%%%%%%%%%%%%%%%%%%%%%%%%%%%%%%%%%%%%%%%%%%%%%%%%%%%%%%%%%%%%

\section{Conclusions}
\label{s:con}

The times of the first close encounter between a pair of planets in equal-mass and evenly spaced systems is strongly affected in the vicinity of first and second-order mean motion resonances between adjacent and next-nearest planets. For planet separations between 4 and 10 mutual Hill radii ($\Delta$), small changes in $\Delta$ can result in variation in close encounter times up to 4 orders of magnitude. The inter-planet spacings of these features are at the nominal locations of period ratios corresponding to first and second-order mean motion resonances between pairs of adjacent planets and next-adjacent planets. There is a sharp increase in the close encounter time at larger separations, which results in systems that can survive at least $10^9$ orbits for spacings of $\Delta \geq 8.5$ and least $10^{10}$ orbits for spacings of $\Delta \geq 8.6$.

The systems investigated in this work had idealised initial conditions with equal mass and evenly-spaced planets in mutual Hill radii on initially circular and co-planar orbits. Further investigations on the role of mean motion resonances in more generalised cases, as well as on developing a detailed understanding of how these resonances affect stability times would be particularly valuable. The stability limits of tightly-packed exoplanet systems of varied stellar mass are particularly interesting due to the upcoming launch of the Transiting Exoplanet Surveying Satellite (TESS), which is scheduled to launch in 2018. The target stars will be F5-M5 and TESS will target stars brighter than the Kepler targets and will be more amenable to radial velocity follow-up \citep{Ric14}. 

Tightly-packed planetary systems display complex dynamical behaviour, and first and second-order MMRs appear to play a large role in their stability. Ultimately, a better understanding of the long-term stability of this class of exoplanet systems is important and needed to study observed systems and those that have yet to be discovered.

%%%%%%%%%%%%%%%%%%%%%%%%%%%%%%%%%%%%%%%%%%%%%%%%%%%%%%%%%%%%%%%%%%%%%%%%%%%%%%%%%%%%%%%%%%%%%%%%%%%
%%%%%%%%%%%%%%%%%%%%%%%%%%%%%%%%%%%%%%%%%%%%%%%%%%%%%%%%%%%%%%%%%%%%%%%%%%%%%%%%%%%%%%%%%%%%%%%%%%%

\section*{Acknowledgements}

We thank the reviewers for their valuable comments and suggestions on this manuscript. We also thank Norm Murray, Cristobal Petrovich, and Billy Quarles for their comments, discussions, and suggestions. A.O additionally thanks Renee Hlozek for her advice and Ari Silburt, Ryan Cloutier, and John Dubinski for help with running the simulations. A.O. acknowledges support from NSERC in the form of a CGSM.

Simulations in this paper made use of the REBOUND code which can be downloaded freely at http://github.com/hannorein/rebound.

\section*{References}

\bibliographystyle{elsarticle-harv} 
\bibliography{bibliography}

\end{document}